\documentclass[cameraready]{Interspeech}
\title{VoxKnesset: A Large-Scale Longitudinal Hebrew Speech Dataset for Aging Speaker Modeling}

\author[affiliation={1},equalcontribution, correspondingauthor]{Yanir}{Marmor}
\author[affiliation={1},equalcontribution]{Arad}{Zulti}
\author[affiliation={1},equalcontribution]{David}{Krongauz}
\author[affiliation={1}]{Adam}{Gabet}
\author[affiliation={2}]{Yoad}{Snapir}
\author[affiliation={2}]{Yair}{Lifshitz}
\author[affiliation={1}]{Eran}{Segal}
\address{
    $^1$ Department of Computer Science and Applied Mathematics, Department of Molecular Cell Biology, Weizmann Institute of Science, Rehovot, Israel \\
    $^2$ ivrit.ai
}
\email{yanir.marmor@weizmann.ac.il}


\usepackage{comment}
\usepackage{subcaption}
\usepackage{adjustbox}

\begin{document}

\maketitle

% the abstract here must exactly match the abstract entered into the paper submission system
\begin{abstract}
    % 1000 characters. ASCII characters only. No citations.
    Speech processing systems face a fundamental challenge: the human voice changes with age, yet few datasets support rigorous longitudinal evaluation. We introduce VoxKnesset, an open-access dataset of ~2,300 hours of Hebrew parliamentary speech spanning 2009–2025, comprising 393 speakers with recording spans of up to 15 years. Each segment includes aligned transcripts and verified demographic metadata from official parliamentary records. We benchmark modern speech embeddings (WavLM-Large, ECAPA-TDNN, Wav2Vec2-XLSR-1B) on age prediction and speaker verification under longitudinal conditions. Speaker verification EER rises from 2.15\% to 4.58\% over 15 years for the strongest model, and cross-sectionally trained age regressors fail to capture within-speaker aging, while longitudinally trained models recover a meaningful temporal signal. We publicly release the dataset and pipeline to support aging-robust speech systems and Hebrew speech processing.
\end{abstract}

\section{Introduction}

From biometric security to automated transcription to health diagnostics, speech processing systems now underpin critical infrastructure across domains — making their vulnerability to the slow, inevitable drift of vocal aging a problem that can no longer be ignored \cite{kelly2013eigenageing, Kelly2016ScoreAgingCF, dmitriev2018locust, snyder2018xvectors,hansen2015speaker}. Natural aging of the vocal folds and vocal tract induces continuous physiological changes that alter acoustic and prosodic patterns over time \cite{decoster2000longitudinal,harrington2007age, rojas2020does, xue2003changes, reubold2010vocal}, degrading speaker verification reliability \cite{jain2016fifty, fairhurst2014selective} and challenging models that rely on stable vocal characteristics, including a growing line of work on automatic age estimation from speech \cite{kelly2012speaker, qin2024investigating, bahari2012age, burkhardt2023speech, alharthi2025tessellated, tursunov2021age, kwasny2021gender}.

%% In order to be at top of second colomn
\begin{figure}[t]
  \centering
  \includegraphics[width=0.85\linewidth]{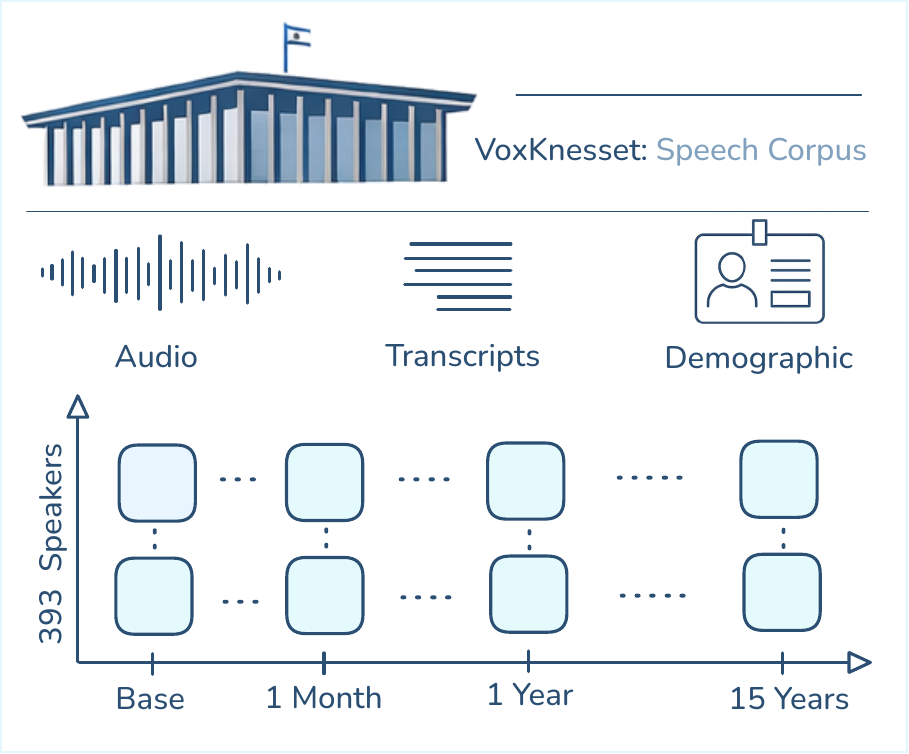}
  \caption{VoxKnesset pairs single-speaker speech segments with aligned transcripts and verified demographic labels across 16 years of parliamentary recordings.}
  \label{imgs:dataset-opening}
\end{figure}

Despite these well-recognized challenges, dataset development has not kept pace. Traditional benchmarks such as TIMIT \cite{garofolo1993timit} provide high-quality recordings with reliable demographic labels but are cross-sectional, capturing each speaker only once. Longitudinal corpora such as CSLU \cite{cole1998cslu}, TCDSA \cite{kelly2012speaker}, MARP \cite{lawson2009multi}, and Greybeard \cite{brandschain2010greybeard} track speakers over time but at the cost of scale, covering small cohorts too sparse for modern deep-learning methods. Large-scale in-the-wild collections such as VoxCeleb2 \cite{chung2018voxceleb2}, even when enriched with estimated age labels (AgeVoxCeleb \cite{tawara2021age}), offer acoustic diversity and scale but were not designed to capture longitudinal vocal change. Recent efforts such as HPP Voice \cite{krongauz2025hpp}, a large Hebrew-speaker cohort with verified clinical metadata, represent progress on label quality and scale, yet with recordings separated by approximately two years, temporal resolution remains limited. The result is a trilemma: no existing resource simultaneously provides (1) dense, repeated recordings of the same speakers spanning many years, (2) sufficient scale and diversity for contemporary models, and (3) trustworthy ground-truth labels.

\begin{figure*}[ht]
    \centering
    \includegraphics[width=\textwidth]{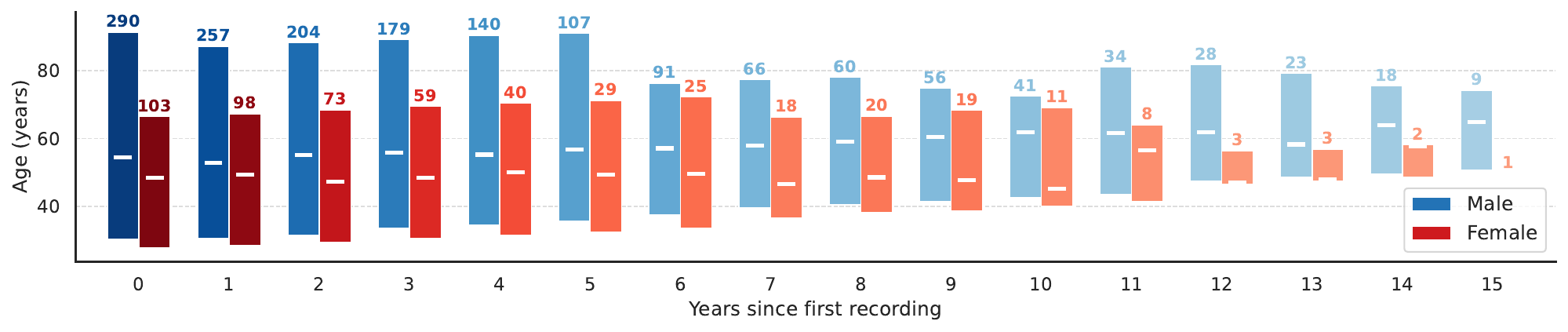}
    \caption{Longitudinal coverage in VoxKnesset by sex. Bars show the age range per recording year; white markers indicate median age; color intensity reflects the number of unique speakers (annotated).}
    \label{fig:dataset_longitudinal_coverage}
\end{figure*}

A recent effort to close this gap is VoxAging \cite{ai2024voxaging}, which provides web-sourced recordings spanning up to 17 years and uses them to benchmark speaker verification under aging conditions. Yet VoxAging also highlights a persistent tension in web-sourced longitudinal data: the difficulty of obtaining verified age labels at scale. In lieu of ground-truth metadata, VoxAging assigns age groups using facial age-recognition algorithms — a practical choice that nonetheless introduces label noise from the recognition model and complicates interpretation of downstream benchmarks, which may partly reflect labeling errors rather than true physiological aging. This underscores the need for longitudinal speech resources grounded in verified, independently sourced demographic metadata, a need the present work directly addresses.

We introduce VoxKnesset, an open-access longitudinal speech dataset derived from 16 years of Israeli parliamentary (Knesset) plenary recordings spanning 2009 to 2025. Parliamentary proceedings offer a unique combination of properties for longitudinal speech research: a consistent, semi-controlled acoustic environment, a large and rotating cast of identified speakers, and rich administrative records that supply verified demographic metadata. VoxKnesset joins a growing family of parliamentary speech resources \cite{erjavec2023parlamint, erjavec2025parlamint, ljubevsic2022parlaspeech, wang2021voxpopuli, virkkunen2023finnish} and is designed explicitly for longitudinal speaker and age analysis. The dataset pairs around 2,300 hours of speaker-attributed speech from 393 Members of Knesset (MKs) with verified ground-truth labels, including age, gender, country of origin, and religion. It aligns each segment with official human-verified session protocols. Many speakers appear repeatedly across multiple parliamentary terms, with time spans of up to 15 years between recordings of the same individual. The dataset is in Hebrew, which is a morphologically rich language that remains relatively underserved in open-access speech data \cite{itai2008language} despite the recent expansion of dedicated resources \cite{marmor2023ivrit, turetzky2024hebdb, sharoni2023saspeech, marmor2025building}. This scarcity makes VoxKnesset a valuable contribution to the broader Hebrew speech processing community in addition to its primary longitudinal focus.

We use this resource to make three contributions:
\begin{enumerate}
\item \textbf{Dataset Release:} We publicly release VoxKnesset, the first large-scale longitudinal Hebrew speech dataset with verified demographic labels and high-quality aligned transcripts.
\item \textbf{Longitudinal Benchmarking:} We evaluate modern speech embeddings such as ECAPA-TDNN \cite{desplanques2020ecapa}, Wav2Vec2-XLSR-1B \cite{baevski2020wav2vec, babu2021xls}, and WavLM \cite{chen2022wavlm} on both age prediction and speaker verification. Specifically, we quantify how performance degrades as the enrollment–test time gap increases.
\item \textbf{Cross-Dataset Age Prediction:} We evaluate age prediction across TIMIT, HPP-Voice, AgeVoxCeleb, and VoxKnesset using modern speech embeddings, showing that VoxKnesset yields results consistent with established benchmarks while introducing cross-lingual and longitudinal dimensions previously unavailable.
\end{enumerate}

\section{Dataset}
\begin{figure}[t]
    \centering
    \includegraphics[width=\linewidth]{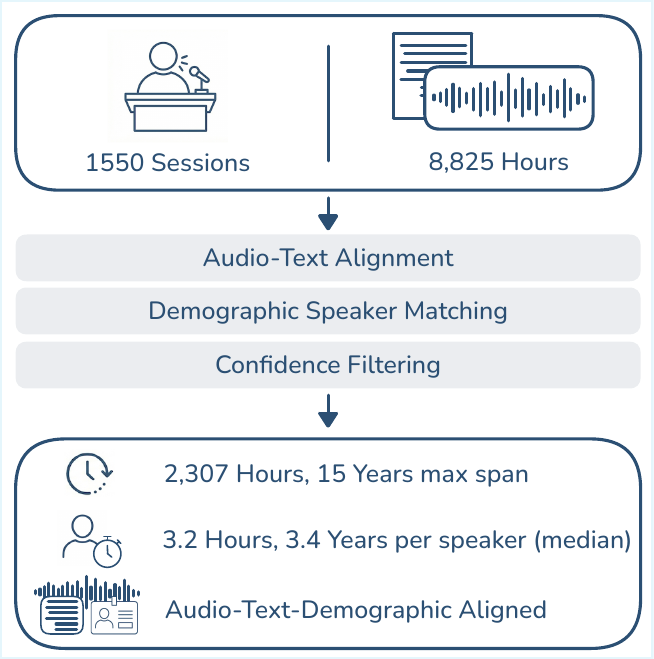}
    \caption{Data curation pipeline: from raw parliamentary recordings to the speaker-attributed longitudinal subset.}
    \label{fig:data-pipeline}
\end{figure} 

\begin{figure}[ht]
    \centering
    \includegraphics[width=\linewidth]{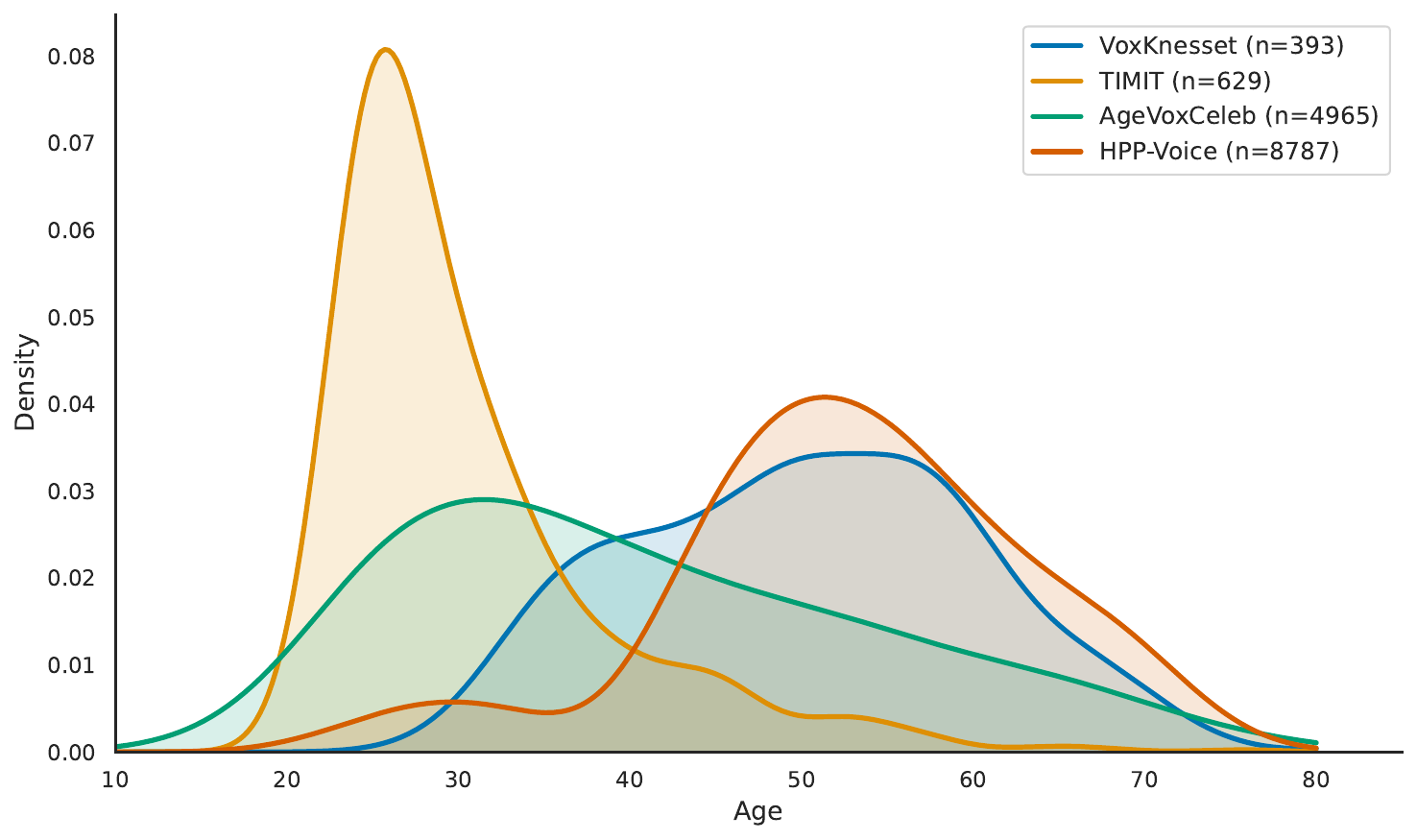}
    \caption{Age distributions across commonly used speech datasets, illustrating differences in demographic coverage.}
    \label{fig:dataset_age_distribution}
\end{figure}

\subsection{Full Corpus}
The dataset is derived from official plenary recordings and transcripts of the Israeli Knesset spanning sixteen years (March 2009 – February 2025), comprising approximately 8,825 hours across roughly 1,550 plenary sessions. Raw parliamentary data pairs audiovisual recordings with timestamped protocol documents; however, these are not directly usable for speech modeling due to timestamp inconsistencies and transcript normalization artifacts.

We implemented a multi-stage alignment pipeline. Audio was extracted from A/V files at 16 kHz mono; timestamps were corrected for backward jumps and drift; and word-level forced alignment was performed using Whisper \cite{radford2023robust} and a Hebrew-adapted variant, following the recovery strategy of Stable-Whisper \cite{stable-ts}. Each word was assigned an alignment confidence score, and segment-level quality was computed as the median of the word probabilities. The corpus reflects real-world parliamentary conditions, including overlapping speech, audience noise, and non-verbatim transcription. Confidence scores are provided so researchers can apply quality thresholds appropriate for their task. The full pipeline and processing code are publicly available.\footnote{Anonymous link}%https://huggingface.co/datasets/ivrit-ai/knesset-plenums}

\subsection{Speaker-Attributed Longitudinal Subset}
To construct a speaker-identified subset, we leveraged the Knesset Corpus \cite{goldin2025knesset}, a curated collection of parsed parliamentary official transcripts enriched with structured demographic metadata for all MKs. We matched transcript segments from the Knesset Corpus against our audio-aligned full corpus, retaining only segments that satisfy three criteria: (1) high audio-alignment confidence; (2) high textual similarity, and (3) a minimum duration of 30 seconds for robust embedding extraction.

The resulting subset comprises 2,307 hours of speaker-attributed speech from 393 Members of Knesset. All speakers carry verified birth year and gender labels drawn from official parliamentary records. The subset captures substantial longitudinal depth: the median age span per speaker is 3.4 years, 47 speakers (12\%) span more than 10 years, and the maximum observed span is 15 years (see Fig. \ref{fig:dataset_longitudinal_coverage}). Data was split 80/20 by speaker, ensuring all sessions from a given speaker appear in the same partition.

\subsection{Demographic Validation}
To validate that VoxKnesset captures meaningful speaker variation, we probe the demographic signal encoded in pretrained speech representations. We extract 1024-dimensional WavLM-Large embeddings (mean-pooled) and fit Ridge classifiers for
gender, religion (4 classes), and birthplace (6 groups), see table \ref{tab:demo_cls}. Gender prediction is near-perfect (99.9\%). Religion and birth-region accuracies (94.8\% and 84.9\%) are close to the majority baselines, reflecting
natural parliamentary class imbalance; however, per-class analysis
reveals a meaningful signal — Muslim speakers are identified with
75.2\% recall, may reflect L1 transfer from Arabic, and former Soviet Union origin speakers with 78.4\% recall, consistent with detectable
Russian accent features. These results suggest that VoxKnesset's demographic metadata may support future research into accent detection and demographic bias auditing in speech systems.

\begin{table}[t]
\centering
\begin{tabular}{lcccc}
\toprule
Task & \#C & AUC & Maj. \\
\midrule
Gender       & 2 & 1.00 & .86 \\
Religion     & 4 & .95  & .95  \\
Birthplace & 6 & .73  & .85  \\
\bottomrule
\end{tabular}
\caption{Demographic prediction with WavLM-L embeddings (Ridge). \#C = classes; Maj. = majority baseline.}
\label{tab:demo_cls}
\end{table}

\section{Experiments and Results}
\label{sec:experiments}

\subsection{Setup}
\label{sec:setup}

All experiments use pretrained speech encoders applied to 10-second
segments. We compare three embedding architectures:
WavLM-Large (1024-d, mean-pooled from the final
transformer layer), ECAPA-TDNN (192-d,
speaker verification embedding), and Wav2Vec2-XLSR-1B (1024-d, mean-pooled). For age
prediction, a Ridge regressor is trained on the resulting
representations. For speaker verification, we additionally evaluate
ERes2Net-Large \cite{chen2023enhanced}, SDPN \cite{chen2025self}, and WavLM fine-tuned for speaker verification. All evaluations are speaker-disjoint.

% ============================================================
\subsection{Cross-Dataset Age Prediction}

We evaluate age prediction both within and across datasets
to assess how well speech embeddings encode chronological age
and whether this signal generalizes across languages and recording conditions.

\subsubsection{Datasets.} Our evaluation spans four corpora with verified age labels
(see Fig.~\ref{fig:dataset_age_distribution} for age distributions):
\textbf{TIMIT}~\cite{garofolo1993timit} (461/168 train/test, ages 21--76, read English),
\textbf{HPP-Voice}~\cite{krongauz2025hpp} (7{,}029/1{,}758, ages 19--81, Hebrew),
\textbf{AgeVoxCeleb}~\cite{tawara2021age} (4{,}468/497, ages 5--95, celebrity interviews), and
\textbf{VoxKnesset} (314/79, ages 28--81, Hebrew parliamentary speech).

\subsubsection{Within-dataset evaluation.}
Table~\ref{tab:within} reports in-domain results.
WavLM-Large consistently achieves the best performance across all corpora,
with MAE ranging from 4.6 years (TIMIT) to 7.3 years (AgeVoxCeleb).
VoxKnesset yields an MAE of 6.3 years --- comparable to the other corpora
despite differing in language and recording domain.
ECAPA-TDNN provides moderate performance, while Wav2Vec2-XLSR-1B trails despite 
its larger parameter count.

\subsubsection{Cross-corpus transfer.}
To test whether the age signal generalizes across corpora, we perform
a leave-one-dataset-out (LODO) evaluation: a Ridge regressor is trained on three
datasets and tested on the held-out fourth (Table~\ref{tab:within}).
VoxKnesset emerges as the most transferable target, achieving
$R^2{=}0.33$ and MAE${=}7.6$ years when held out --- the smallest
domain gap ($\Delta R^2{=}0.09$) of any corpus.

\setlength{\tabcolsep}{1.5pt}
\begin{table}[t]
\small
\caption{Age prediction ($R^2$/MAE in years) within-domain and under leave-one-dataset-out
(LODO) cross-corpus transfer, using WavLM-Large for transfer.
$\Delta R^2$ measures the in-domain vs.\ transfer gap.}
\label{tab:within}
\centering
\begin{tabular}{l ccc c c}
\toprule
& \multicolumn{3}{c}{\textbf{Within-dataset }} 
& \textbf{LODO} 
&  \\
\cmidrule(lr){2-4}
Dataset 
& WavLM 
& ECAPA 
& XLS-R 
& WavLM
&  \boldmath$\Delta R^2$ \\
\midrule
VoxKnesset  & \textbf{0.42 / 6.3} & 0.24 / 7.2 & 0.24 / 7.5 & \underline{0.33 / 7.6} & \underline{0.09} \\
TIMIT       & \textbf{0.43 / 4.6} & 0.27 / 5.2 & 0.14 / 5.6 & $-$2.52 / 14.2      & 2.95 \\
AgeVoxCeleb & \textbf{0.57 / 7.3} & 0.42 / 8.6 & 0.29 / 9.5 & 0.17 / 10.2         & 0.40 \\
HPP-Voice   & \textbf{0.56 / 5.4} & 0.38 / 6.5 & 0.35 / 6.5 & 0.01 / 8.7          & 0.55 \\
\bottomrule
\end{tabular}
\end{table}

% ============================================================
\subsection{Longitudinal Analysis}
\label{sec:longitudinal}

Having established a cross-sectional age signal in speech embeddings, we leverage VoxKnesset's longitudinal structure to ask three progressively deeper questions: whether embeddings drift with speaker age, whether that drift encodes aging, and whether it degrades speaker verification.

\begin{figure}[ht]
    \centering

    % (a) UMAP
    \begin{subfigure}{0.95\linewidth}
        \centering
        \includegraphics[width=\linewidth]{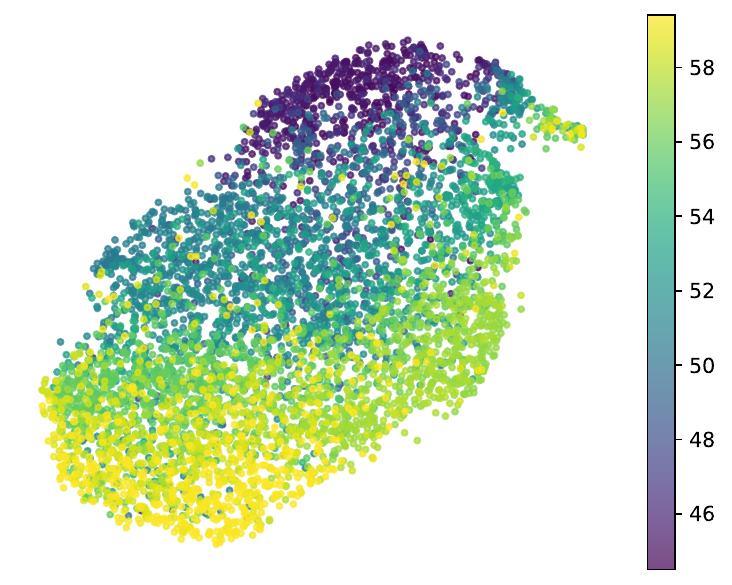}
        \caption{UMAP visualization of speaker embeddings by age.}
        \label{fig:umap}
    \end{subfigure}
    
    % \vspace{1em}
    
    % (b) Age prediction
    \begin{subfigure}{0.95\linewidth}
        \centering
        \includegraphics[width=\linewidth]{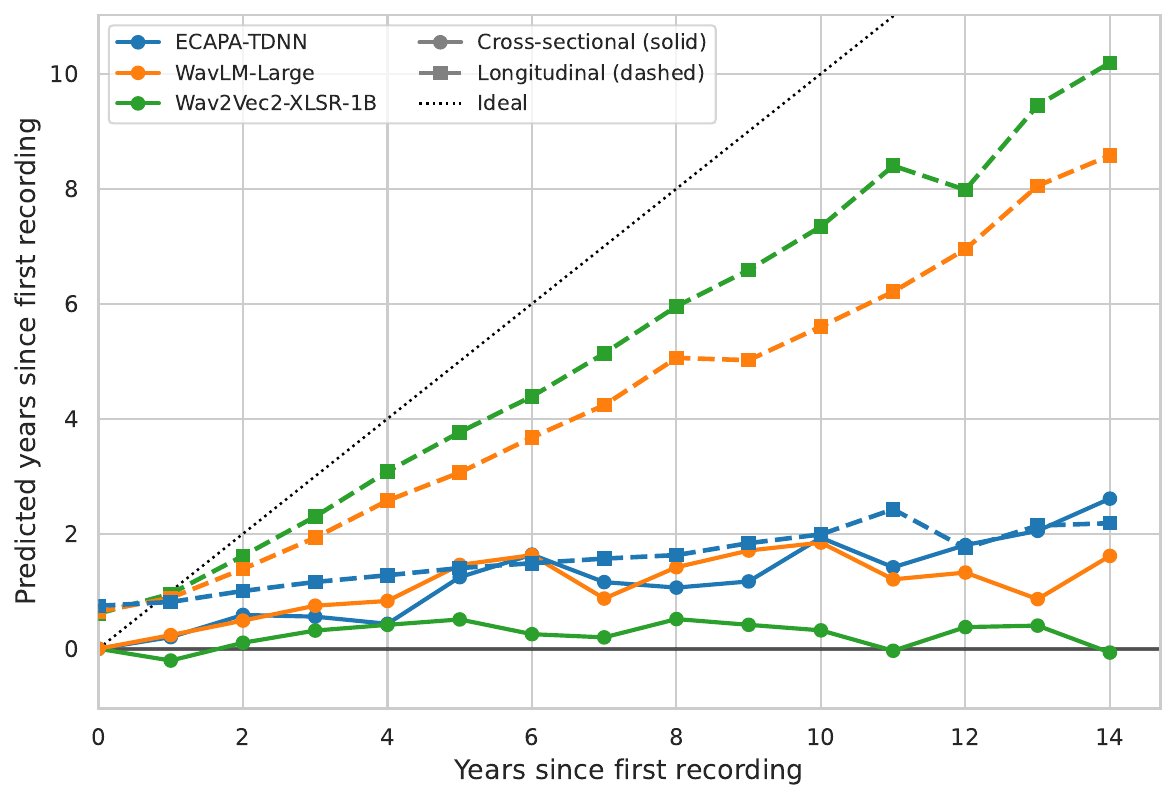}
        \caption{Predicted age delta as a function of recording span.}
        \label{fig:age_delta}
    \end{subfigure}
    
    % \vspace{1em}
    
    % (c) Speaker verification EER
    \begin{subfigure}{0.95\linewidth}
        \centering
        \includegraphics[width=\linewidth]{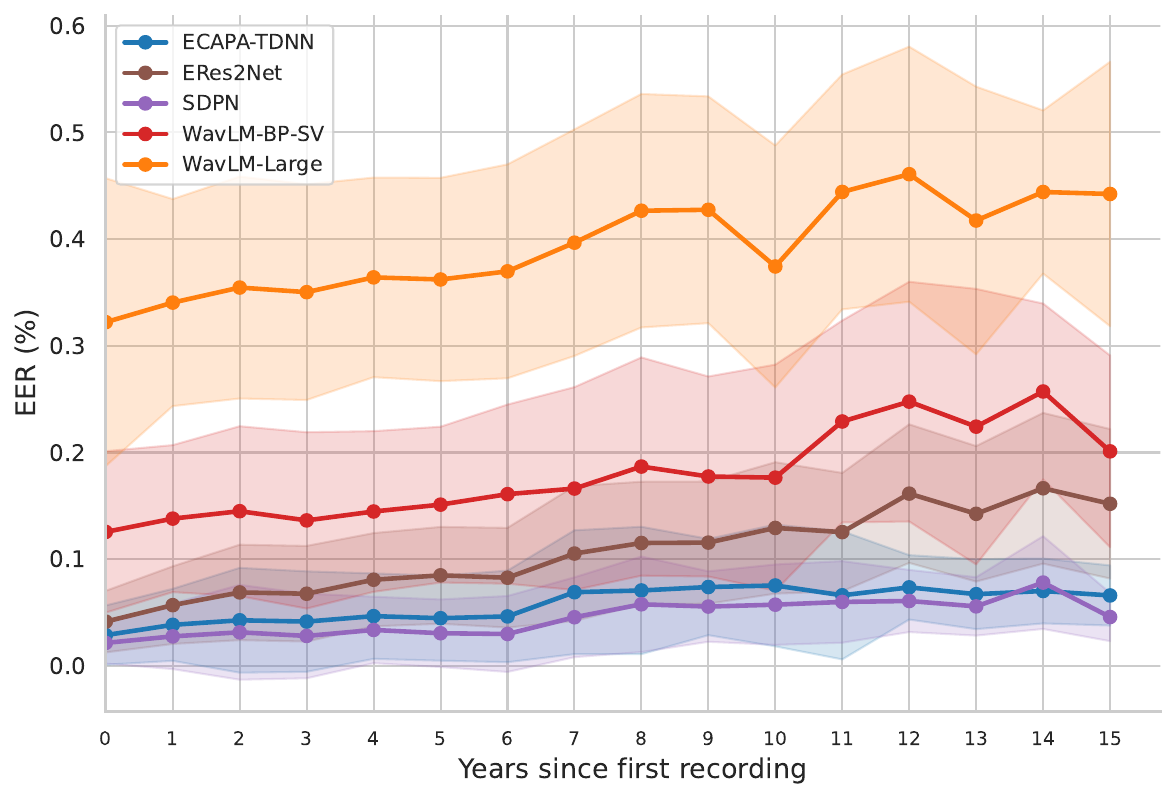}
        \caption{Speaker verification equal error rate (EER) over time.}
        \label{fig:speaker_verification}
    \end{subfigure}
    
    \caption{Longitudinal effects of speaker aging: (a) Embedding space distribution, (b) age prediction accuracy, and (c) verification performance degradation.}
    \label{fig:longitudinal_results}
\end{figure}

\textbf{Do embeddings drift over time?}
Fig.~\ref{fig:umap} illustrates this for a representative speaker, whose embeddings trace a clear age gradient in UMAP space. The subsequent analyses confirm that this temporal structure holds at scale in both age estimation (Fig.~\ref{fig:age_delta}) and verification degradation (Fig.~\ref{fig:speaker_verification}).

\textbf{Does the drift encode aging?}
We evaluate two paradigms (Fig.~\ref{fig:age_delta}). In the \emph{cross-sectional transfer} setting, a Ridge regressor trained on absolute age is applied longitudinally by computing the predicted age difference between follow-up and baseline recordings (solid lines). Predicted deltas plateau at 1--2 years regardless of true elapsed time, indicating that cross-sectional regressors capture between-speaker variation and do not generalize to within-speaker change. In the \emph{longitudinal} setting, an MLP trained on concatenated embedding pairs from different years of the same speaker predicts elapsed time directly (dashed lines). WavLM-Large and Wav2Vec2-XLSR-1B scale roughly monotonically with true elapsed time, reaching ${\sim}$8.5 and ${\sim}$10 predicted years at $\Delta{=}14$, with increasing underestimation at longer gaps. ECAPA-TDNN remains flat under both paradigms, indicating minimal encoded aging information. By contrast, the strong longitudinal performance of Wav2Vec2-XLSR-1B suggests that aging information occupies a recoverable subspace of its embeddings.

\textbf{Does aging degrade speaker verification?}
Finally, we quantify the practical consequences of vocal aging for deployed systems. Each speaker's earliest recorded year serves as the enrollment baseline. For each time gap $\Delta \in \{0, \ldots, 15\}$~years, the evaluation cohort comprises all same-sex recordings at the target age (baseline age~$+ \Delta$); speakers are excluded from a given $\Delta$ if no recordings exist at that age or if they contribute more than 20\% of cohort duration. Results are reported in Fig.~\ref{fig:speaker_verification}.

\section{Conclusion}
We introduced VoxKnesset, a 2,300-hour longitudinal Hebrew speech dataset
spanning up to 15 years per speaker with verified demographic labels and
aligned transcripts. Our experiments reveal three key findings: the age
signal in general-purpose embeddings transfers across languages and corpora;
speaker-verification EER more than doubles over a 15-year gap even for the
strongest model; and models trained on paired longitudinal embeddings recover
within-speaker aging signal that cross-sectional regressors miss entirely.

Several limitations should be noted. VoxKnesset captures a single speech
register (parliamentary debate) with a demographic skew toward older adults,
and recording conditions likely evolved over the 16 years. Disentangling
channel drift from biological aging remains an open challenge.

Looking ahead, VoxKnesset's per-speaker longitudinal recordings paired with
aligned transcripts could support aging-aware re-enrollment strategies for
biometric systems, speaker-adaptive voice technologies that track individual vocal change
over time. We publicly release the
dataset and processing pipeline to support these directions.

\clearpage

\bibliographystyle{IEEEtran}
\bibliography{mybib}

\end{document}